\documentclass[%
reprint,
 amsmath,amssymb,
 aps,
]{revtex4-1}

\usepackage{graphicx}% Include figure files
\usepackage{dcolumn}% Align table columns on decimal point
\usepackage{bm}% bold math
\usepackage{float}
\usepackage[T1]{fontenc}
\usepackage[latin9]{inputenc}
\usepackage{amssymb}
\usepackage{color}
\usepackage{graphicx,color}
\usepackage{amsmath, amssymb, graphics}
\usepackage{qcircuit}
\usepackage{pgf, tikz}
\usetikzlibrary{arrows, automata}
\makeatletter
\numberwithin{equation}{section} %% Comment out for sequentially-numbered
  \@ifundefined{theoremstyle}{\usepackage{amsthm}}{}
  \theoremstyle{plain}
  \newtheorem{thm}{Theorem}[section]
  \theoremstyle{plain}
  
  \theoremstyle{plain}
  
  \theoremstyle{remark}
  
  \theoremstyle{remark}
  
  \theoremstyle{plain}

\tikzset{
circleA/.style={
  circle,
  inner sep=0pt,
  text width=6mm,
  align=center,
  draw=black,
  fill=white
  }
}

\tikzset{
circleB/.style={
  circle,
  inner sep=0pt,
  text width=8mm,
  align=center,
  draw=black,
  fill=white
  }
}

\tikzset{
circleC/.style={
  circle,
  inner sep=0pt,
  text width=11mm,
  align=center,
  draw=black,
  fill=white
  }
}

\smallskip
\def\<{{\langle }}
\def\>{{\rangle }}

\def\ket#1{|#1\rangle}

\smallskip
\def\<{{\langle }}
\def\>{{\rangle }}

\def\bar#1{\overline{#1}}
\def\ha#1{\hat{#1}}
\def\ti#1{\tilde{#1}}

\begin{document}

%\preprint{APS/123-QED}

\title{Five-qubit states generated by Clifford gates}% Force line breaks with \\
%\thanks{A footnote to the article title}%

%\altaffiliation[Also at ]{Central Connecticut State University}%Lines break automatically or can be forced with \\
\author{Fr\'ed\'eric Latour}%
 \email{latourfre@ccsu.edu}
 \author{Oscar Perdomo}
 \email{perdomoosm@ccsu.edu}
 \thanks{Corresponding author}
\affiliation{
Central Connecticut State University 
%\textbackslash\textbackslash
}
%

%\collaboration{MUSO Collaboration}%\noaffiliation

%\collaboration{CLEO Collaboration}%\noaffiliation

\date{\today}% It is always \today, today,
             %  but any date may be explicitly specified

\begin{abstract}  The Clifford group is the set of gates generated by controlled-$Z$ (CZ) gates and the two local gates $P=\begin{pmatrix} 1&0\\0&i\end{pmatrix}$ and  $H=\frac{1}{\sqrt{2}} \begin{pmatrix} 1&1\\1&-1\end{pmatrix}$. We will say that a $n$-qubit state is a Clifford state if it can be prepared using Clifford gates, this is, $\ket{\phi}$ is Clifford if 
$\ket{\phi}=U\ket{0\dots 0}$ where $U$ is a Clifford gate. These states are known as the stabilizer states and they arise in quantum error correction. 
In this paper we study the set of all $5$-qubit Clifford states. By using an exhaustive method we confirm  that there are $19388160$ states. We point out that, if we measure entanglement entropy as the average of the von Neumann entropy of the reduced density matrices obtained by considering all possible subsets of possible two-qubits, then the possible entanglement entropies reached by  these Clifford states are $0,\frac{3}{5},\frac{9}{10},1,\frac{6}{5},\frac{7}{5},\frac{3}{2},\frac{8}{5},\frac{9}{5}$ and $2$. The main goal of the paper is to understand the action of the CZ gates action on the 5-qubit states. With this goal in mind,  we partition the Clifford states into orbits using the equivalence relation: two states are equivalent if they differ by a local Clifford gate. We show that there are 93 orbits, and we label each orbit in such a way that it is easy to see the effect of the CZ gates. Diagrams and tables explaining the action of the CZ gates on all the orbits are presented in the paper. A similar work is done for  the real Clifford 5-qubits states, this is, for states that can be prepared with CZ gates, the $Z=\begin{pmatrix} 1&0\\0&-1\end{pmatrix}$ and the Hadamard gate.

%Finally we show that the set of populations of these states include all 16 possibilities of the form $\ket{i_1i_2i_3i_4}$, it includes all 120 possibilities of the form $\frac{1}{\sqrt{2}}\ket{i_1i_2i_3i_4}+\frac{1}{\sqrt{2}}\ket{j_1j_2j_3j_4}$, in other words, it includes all 120 possibilities where only two elements in the canonical basis show up with equal probability $\frac{1}{2}$. It include only 140 possibilities where exactly four elements in the canonical basis shows up with equal probability $\frac{1}{4}$. It include only 30 possibilities where exactly 8 elements in the canonical basis shows up with equal probability and finally it include the case where all the 16 elements in the canonical base show up with equal probability. Overall, only 307 possible populations are possible. 
\end{abstract}
%\begin{description}
%\item[Usage]
%Secondary publications and information retrieval purposes.
%\item[PACS numbers]
%May be entered using the \verb+\pacs{#1}+ command.
%\item[Structure]
%You may use the \texttt{description} environment to structure your abstract;
%use the optional argument of the \verb+\item+ command to give the category of each item. 
%\end{description}

%\pacs{Valid PACS appear here}% PACS, the Physics and Astronomy
                             % Classification Scheme.
%\keywords{Suggested keywords}%Use showkeys class option if keyword
                              %display desired
\maketitle

%\tableofcontents

%$\ket{\phi_1}$ and $\ket{\phi_2}$,

\onecolumngrid %makes it one column

\section{Introduction} Entanglement is one of the most appealing properties in quantum information. This paper studies how entanglement occurs for 5-qubit states when we limit ourselves to gates in the Clifford group. More precisely, we study the set of pure $5$-qubit states that can be prepared using gates in the Clifford group. We call these states, {\em Clifford states}. Notice that another way to describe the Clifford states is as the set of possible first columns of matrices in the Clifford group. More than describing the Clifford $5$-qubit states, we provide efficient circuits---with a minimum number of non-local gates---that prepare them, and we split the Clifford states into orbits (or equivalence classes) with the property that states in each orbit differ by a local Clifford gate. This is, two states $\ket{\phi_1}$ and $\ket{\phi_2}$ are in the same orbit if $\ket{\phi_2}=(U_1\otimes\dots \otimes U_5)\ket{\phi_1},$ where $U_i$ is generated by the Hadamard gate and the phase gate $P=\begin{pmatrix} 1&0\\0&i\end{pmatrix}$. Notice that, in particular, states in the same orbit have the same entanglement entropy. After thus partitioning the Clifford states, we explain how to ``navigate'' these orbits using controlled-$Z$ gates, we do this by providing tables with information regarding how from each orbit we can jump into another using CZ gates. This study provides a better understanding of the Clifford group. Paper \cite{T} and the references within has more details regarding the importance of the Clifford group. In order to describe the Clifford states as stabilizer states, we need to consider the group $P_n$ of $n$-qubit Pauli operators as those matrices of the form $cA_1\otimes\dots\otimes A_n$ where $c\in \{1,-1,i,-i\}$ and $A_i\in \{I,X,Y,Z\},$  with $I=\begin{pmatrix} 1&0\\0&1\end{pmatrix}$, $X=\begin{pmatrix} 0&1\\1&0\end{pmatrix}$, $Y=\begin{pmatrix} 0&-i\\i&0\end{pmatrix}$ and  $Z=\begin{pmatrix} 1&0\\0&-1\end{pmatrix}$. Clearly $P_n $ has $4^{n+1}$ elements. A quantum state $\ket{\psi}$ stabilizes a unitary matrix $U$ if $U\ket{\psi}=\ket{\psi}$ (in this definition we do not ignore the global phase of $\ket{\psi}$). The Clifford states are stabilizer states because they are characterized as those states that are stabilized by exactly $2^n$ matrices in $P_n$. See Theorem 1 in \cite{AG}. For a more detailed approach to stabilizer states and their relation with quantum error correction and fault-tolerant quantum computation, we refer to Chapter 10 in \cite{NC}.

The authors have been also interested \cite{LP}, \cite{P} in studying what they  called, ``real Clifford states'' which are Clifford states with real amplitudes or equivalently, states that can be prepared with the Hadamard gate, the Z-gate with $Z=\begin{pmatrix} 1&0\\0&-1\end{pmatrix},$ and CZ gates. We partition the set of real Clifford states, but this time, we organize them in orbits with the property two states $\ket{\phi_1}$ and $\ket{\phi_2}$ are in the same orbit if $\ket{\phi_2}=(U_1\otimes\dots \otimes U_5)\ket{\phi_1}$ with $U_i$ is generated by the Hadamard gate and the Z-gate.

The main results in this paper are obtained by exhaustion and form a continuation of the work done in \cite{P} for three Clifford 3-qubit states, and in \cite{LP} for Clifford 4-qubit states. We have noticed that the number or Clifford states is considerably larger than the number of real Clifford states. For 5-qubit states, while there are nearly $20$ million Clifford 5-qubit states, $19388160$ to be exact, we only have nearly $300$ thousand real Clifford states, $293760$ to be exact. The number of $n$-qubit Clifford states can be counted using a beautiful algebraic argument, using the fact that they are stabilizer states. We can count the number of subgroups of order $2^n$ of $P_n$ that define stabilizer states, and it turns out that each one of these subgroups must be generated by $n$ commuting, linearly independent elements. By counting the generators of each subgroup, it can be deduced, see Proposition 1 in \cite{AG}, that the number of Clifford $n$-qubit states (ignoring the global phase) is  $2^n\Pi_{k=0}^{n-1} (2^{n-k}+1)$, which is consistent with our results for $n\leq5.$ We are hoping that our work can lead to a different way of counting the Clifford $n$-qubit states, exploiting the way that they are generated by controlled-$Z$ and local  Clifford gates. We have also noticed the following:

{\bf Conjecture:} The number of real $(n+1)$-qubit Clifford states equals the number of $n$-qubit Clifford states.

So far we only have a proof for $n=2,3,4,$ and $5$.

As a bonus from our study, we get an expression for all the absolutely maximally entangled states, AME states, that can be prepared with Clifford gates. We have that there are 1990656 Clifford AME 5-qubit states. AME states are states with the property that all reduced states obtained by tracing out at least half of the particles are maximally mixed \cite{H1}. We already knew that some AME can be generated using only the Hadamard gate and CZ gates, see for example \cite{C}. The representation found in \cite{C} is not only is given in terms of Hadamard gates but also it is given by a graph. See \cite{He} for the definition of graph states. We notice that, even though the number of graph states is relatively small compared with the number of Clifford states, every orbit contains a graph state. Section \ref{graphs} shows the circuits representing these graph states. Regarding this last observation, M. Grassl et al. showed in \cite{G} that every $n$-qubit Clifford state is equivalent to a state that can be represented with a graph. In particular, we have that every orbit contains a state that  can be prepared using only CZ and Hadamard gates. Other partitions of the set of Clifford states have been considered in \cite{KMP}. 
  
An important result that makes the study of Clifford states worthwhile is the Gottesman-Knill theorem (see Theorem 10.7 in \cite{NC}):
 
 \begin{thm} Suppose a quantum computation is performed which involves only the following elements: state preparation in the computation basis, Hadamard gates, phase gates, controlled-Z gates (or controlled-NOT gates which are equivalent), Pauli gates, and measurements of observables in the Pauli group, together with the possibility of classical control conditioned on the outcome of such measurements. Such a computation may be efficiently simulated on a classical computer.
 \end{thm}
 
The states that can be prepared in the theorem above are the Clifford states. In this way the results in this paper help us to understand the gap between classical computers and quantum computers.
For example, even though there is a large variety of entanglement types that can be reached among the Clifford states, it seems that, in order to prepare every Clifford state using local Clifford gates and controlled-$Z$ gates, the number of controlled-$Z$ gates that are needed is relatively small. On the one hand, when we look at the full set of all qubit states (not just the Clifford states), every $2$-qubit state can be prepared using local gates and one controlled-$Z$ gate. For $n=3,$ we need three controlled-$Z$ gates to prepare every 3-qubit state, and it is known that this result cannot be improved (that is, the statement would be false if ``three'' were changed to ``two'') \cite{Z}. For $n=4$ the minimum number of controlled-$Z$ gates needed to prepare any every 4-qubit state is unknown. We know that at least six are needed \cite{PI}, but the ``best'' circuit that is known to prepare every 4-qubit state has 8 controlled-$Z$ gates \cite{CK}.
For $n=5,$ at least 13 controlled-$Z$ gates are needed \cite{PI}; however, the ``best'' circuit (in the sense of ``fewest controlled-$Z$ gates'') that is known to prepare every 5-qubit state uses 26 controlled-$Z$ gates \cite{CK}. We compare this to the Clifford case: in order to prepare every Clifford state using local Clifford gates and controlled-$Z$ gates, the minimum number of controlled-$Z$ gates needed is 1 for $n=2$, 2 for $n=3$, 3 for $n=4$, and 5 for $n=5$. But we do not know the answer for higher values of $n.$ Thus, an open problem in this direction that will measure the complexity of the Clifford states is the following:

{\bf Open Problem} {\it Find the minimum number of controlled-$Z$ gates needed to prepare every $n$-qubit Clifford state, using only local Clifford gates and controlled-$Z$ gates.}

An interesting observation is that the answer for the open problem above may be different if we change ``local Clifford gates'' to just ``local gates'' and this is due to the LU-LC conjecture that was listed at the 28th open problem in quantum information \cite{LULC}. This conjecture states that if two Clifford states can be connected with a local unitary matrix, that is, if $\ket{\phi_1}=U\ket{\phi_2}$, then there exists a local Clifford gate $V$ such that $\ket{\phi_1}=V\ket{\phi_2}$. This conjecture was proven to be false by explicit construction of a counterexample. There are two $27$-qubit Clifford states that could not be connected with a local Clifford gate but they could be connected with a local unitary matrix, see \cite{JC}.

We can also consider the same problem for real $n$-qubit Clifford states, using local real Clifford gates and controlled-$Z$ gates. In this case, the minimum number of controlled $Z$-gates needed is 1 for $n=2$, 3 for $n=3$, 5 for $n=4$, and 7 for $n=5$.
  
{\bf Open Problem} {\it Find the minimum number of controlled-$Z$ gates needed to prepare every $n$-qubit real Clifford state, using only real local Clifford gates and controlled-$Z$ gates.}

There are several reasons to study real Clifford states; one of them is that, because their orbits are significantly smaller than those of all Clifford states, they can be used as a starting point to easily construct all Clifford states. Another reason is that the structure of the orbits for real Clifford states can be much more intricate and interesting than that for all Clifford states. For example, as we will see in Section IV, the largest orbit for all 5-qubit Clifford states (the one containing the AME states) contains twelve real Clifford orbits, and these twelve orbits can be labeled using bidirectional cycles in a way that allows prediction of the result of the action of the controlled-$Z$ gates. This did not happen for 4-qubit Clifford states, and thus one wonders if similar behavior is exhibited by the largest orbit for the $n \geq 6$ case.
 
\section{Main results for Clifford states}

The results in this paper were found by exhaustion. This is the method that we used: we took the state $\ket{00000}$ and then we considered all the states of the form $A\ket{000000}$ when $A$ is a local Clifford gate. These 62208 states constitute the orbit of the Clifford state $\ket{000000};$ we called this orbit $S_0,$ and we placed it at the top of our orbit connectivity graph (``level zero''). Recall that we are not considering two states to be the same if they differ by a global phase; if we were to identify such states, the size of the orbit would have been $62208/8= 7776.$ Once we had $S_0$, we considered the set $CZ(1,2)S_0=\{CZ(1,2) A: A\in S_0 \}$, where $CZ(i,j)$ denotes the controlled-$Z$ gate acting on qubits $i$ and $j$. We are labeling the qubits 1 through 5, from left to right. We then considered the set difference $CZ(1,2)S_0\setminus S_0$ and we selected a Clifford state $\ket{\phi_1}$ in this set. We continued by computing the orbit of this state  $\{A\ket{\phi_1}: \hbox{$A$ is a local Clifford gate } \}$  and we called this orbit $T_{12},$ which we placed lower than $S_0$ on our connectivity graph (``level one''), and connected it to $S_0$ by an edge. If necessary, we would have then performed the same computation with an element of the set difference $CZ(1,2)S_0\setminus (S_0 \cup T_{1,2})$ to get another orbit; however, that set difference turned out to be empty. We performed the same computation with the other 9 sets $CZ(i,j)S_0$ to obtain 9 other orbits $T_{ij},$ also at level one and connected to $S_0.$ We continued this process with the orbit $S_0$ replaced by each of the $T_{ij};$ creating new orbits at level two of the graph, and so on. The process ended when reaching a level at which, for every orbit $O,$ the sets $CZ(i,j)O$ were all subsets of the union of the orbits that had already been found.

We thus proved that there are 93 orbits. They are distributed as follows: one orbit $S_0$,  10 orbits labelled $T_{ij}$, 10 orbits labelled $U_{ijk}$, 15 orbits labelled $U_{ij/kl}$, 5 orbits of type $V_{ijkl}$, 15 orbits of type $V_{ij/kl}$, 10 orbits of type $V_{ij}$, one orbit of type $W_0$, 10 orbits of type $W_{ij}$, 15 orbits of type $W_{ij/kl},$ and one orbit of type $X_0$. Two orbits are considered to be of the same type if they are identical up to permutations of the five qubits.

The letters designating the orbits were chosen based on the minimum number of CZ gates needed in order to reach them starting from $S_0.$ Orbits labeled with the letter $T$ require one CZ gate, those labeled with the letter $U$ require two, those labeled with the letter $V$ require three, those labeled with the letter $W$ require four, and the one labeled with the letter $X$ requires five, which is the largest minimum number of CZ gates needed to reach any Clifford state.

In the name of any orbit, there cannot be any repeated numbers; in other words, $i, j, k, l$ represent four distinct numbers, which are always elements of the set $\{1, 2, 3, 4, 5\}$ (except in the three cases where the is a single orbit with the subscript $0).$
The order in which the individual numbers are listed in an orbit's name is irrelevant; for instance, $T_{123} = T_{231}.$ In addition, the two numbers before a slash can be exchanged with the two numbers after the slash; for instance, $W_{12/35} = W_{35/12}.$

\begin{table}[h]
\setlength\extrarowheight{1mm}
\begin{tabular}{|c|c|c|} %the | are vertical bars, not lowercase Ls
\hline
class type & size & entropy \\
\hline
$S_0$ & $2^8 3^5 = 62208$ & $0$ \\
\hline 
$T_{ij}$ & $2^9 3^4 = 41472$ & $3/5$ \\
\hline 
$U_{ijk}$ & $2^9 3^5 = 124416$ & $9/10$ \\
\hline
$U_{ij/kl}$ & $2^{10} 3^3 = 27648$ & $6/5$ \\
\hline 
$V_{ijkl}$ & $2^9 3^5 = 124416$ & $1$ \\
\hline
$V_{ij/kl}$ & $2^{10} 3^5 =248832$ & $7/5$ \\
\hline 
$V_{ij}$ & $2^{10} 3^4 = 82944$ & $3/2$ \\
\hline 
$W_0$ & $2^9 3^5 = 124416$ & $1$ \\
\hline 
$W_{ij}$ & $2^{10} 3^5 =248832$ & $8/5$ \\
\hline 
$W_{ij/kl}$ & $2^{11} 3^5 =497664$ & $9/5$ \\
\hline 
$X_0$ & $2^{13} 3^5 =1990656$ & $2$ \\
\hline
\end{tabular}
\caption{Size and entropy of the 11 types of orbits for Clifford states.} \label{table1}
\end{table}

Table \ref{table1} shows the size of each orbit, as well as the entanglement entropy of its elements.

%In the notation above, we will assume that anytime a collections of variables is display in an orbit or a statement then, these variables are assumed to be different. For example the $i,j,k$ in $T_{ijk}$ are assume to be different. 
In order to understand the relationships between the different orbits, it is useful to know how each of the CZ gates acts on the elements of each orbit. This information is summarized in Tables \ref{table2} and \ref{table3}, which due to their sizes are included in the appendix.
In these tables, within each row, two different letters always represent two different labels. For instance, in the row corresponding to ``${il}$ acting on $U_{ijk}$'', $l$ represents a label that is distinct from $i, j,$ and $k.$ Thus, the row can be used for the action of $\mathrm{CZ}(1,4)$ on $U_{123}$ or the action of $\mathrm{CZ}(2,5)$ on $U_{124}$, but not for the action of $\mathrm{CZ}(1,4)$ on $U_{145}.$ In addition, the overscore denotes the complement with respect to $\{1,2,3,4,5\};$ for instance, $U_{\bar{34}} = U_{125}.$
%If we write ``${il}$ acting on $T_{ijk}$'' we assume that $i,j,k$ and $l$ are different. We also have that the variables together in an orbit are indistinguishable. Two examples of this property are: (i) the orbits $T_{123}$ and $T_{312}$ and (ii) In a statement of the form `` ${14}$ acting on $T_{123}$ ...'' we can change the $1$ with either  $2$ or $3$ and the statement remains true. We also have that groups in the name of an orbit with the same amount of elements are indistinguishable. For example $T_{13,24}$ is the same as $T_{24,13}$ or $X_{1,23,45}$ is the same as $X_{1,45,23}$.

%The results of applying each controlled Z-gate on the elements of each orbit are shown in Table \ref{table2}:

\begin{figure}
\begin{center}
    \begin{tikzpicture}
        \node at (0,0) [circleA] (A) {$S_0$};
        \node at (0,-1.25) [circleA] (B) {$T_{ij}$};
        \node at (-1,-2.5) [circleA] (C) {$U_{ijk}$};
        \node at (1,-2.5) [circleA] (D) {$U_{\frac{ij}{kl}}$};
        \node at (-2,-3.75) [circleB] (E) {$V_{ijkl}$};
        \node at (0,-3.75) [circleA] (F) {$V_{\frac{ij}{kl}}$};
        \node at (2,-3.75) [circleA] (G) {$V_{ij}$};
        \node at (-2,-5) [circleA] (H) {$W_0$};
        \node at (0,-5) [circleA] (I) {$W_{ij}$};
        \node at (2,-5) [circleA] (J) {$W_{\frac{ij}{kl}}$};
        \node at (0,-6.25) [circleA] (K) {$X_0$};
        \draw (A) -- (B);
        \draw (B) -- (C); 
        \draw (B) -- (D);
        \draw (C) -- (E); 
        \draw (C) -- (F);
        \draw (C) -- (G);
        \draw (D) -- (F); 
        \draw (D) -- (G);
        \draw (E) -- (H);
                \draw (E) -- (F);  
        \draw (E) -- (I);
        \draw (F) -- (I); 
        \draw (F) -- (J);
        \draw (G) -- (I); 
                \draw (H) -- (I); 
                                \draw (J) -- (I); 
        \draw (G) -- (J);
        \draw (J) -- (K); 
    \end{tikzpicture}
    \end{center}
    \caption{Actions of CZ gates on orbits of Clifford states.}  \label{fig1}
    \end{figure}
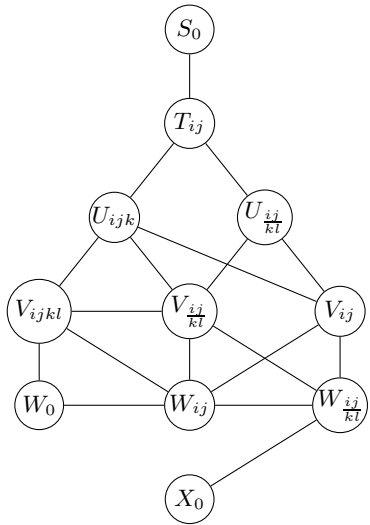

The graph in Figure \ref{fig1} shows the relationship between the 93 orbits; however, in order to make the graph readable, only one of the orbits of each type is displayed. The presence of an edge between two nodes indicates the existence of a CZ gate and of two orbits, one of each type, taking a state in one orbit to a state in the other orbit. Since CZ gates are involutory, all edges are bidirectional.

\section{Circuits preparing states in the Clifford group}\label{graphs} 

We show the circuits representing graph states:
\begin{itemize}
\item
the orbit $S_0$, consisting in all unentangled Clifford states. An element in this orbit is $\ket{00000}$.
$$
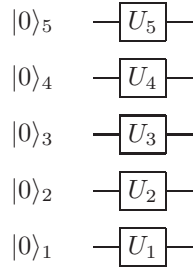
\Qcircuit @C=1em @R=.7em {  \lstick{\ket{0}_5}  & & \gate{U_5} & \qw  \\
\lstick{\ket{0}_4}  & & \gate{U_4} & \qw  \\
\lstick{\ket{0}_3}  & & \gate{U_3} & \qw  \\
\lstick{\ket{0}_2}  & & \gate{U_2} & \qw  \\
\lstick{\ket{0}_1}  & & \gate{U_1} & \qw  \\
                      }$$

Notice that every element in $S_0$ can be prepared with the circuit above, with each $U_i$ a product of Hadamard and phase gates.

\item
A state in $T_{ij}$ can be obtained by applying Hadamard gates to the qubits $i$ and $j$ and then a CZ gate to these two qubits.

$$
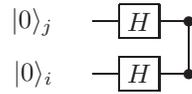
\Qcircuit @C=1em @R=.7em { 
 \lstick{\ket{0}_j}  & & \gate{H} & \ctrl{1}   \\
\lstick{\ket{0}_i}  & & \gate{H} & \ctrl{-1}&
                      }$$
\item
we can reach a state in $U_{ijk}$ as follows,

$$
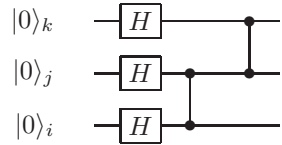
\Qcircuit @C=1em @R=.7em {  \lstick{\ket{0}_k}  & & \gate{H} & \qw & \qw &\ctrl{1}& \qw \\
 \lstick{\ket{0}_j}  & & \gate{H} & \ctrl{1}  & \qw  &\ctrl{-1} & \qw\\
\lstick{\ket{0}_i}  & & \gate{H} & \ctrl{-1} & \qw  & \qw&\qw
                      }$$
\item

we can reach a state in $U_{ij/kl}$ as follows,

$$
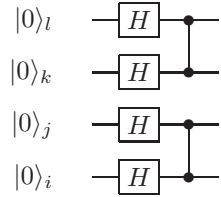
\Qcircuit @C=1em @R=.7em {  \lstick{\ket{0}_l}  & & \gate{H} & \ctrl{1}& \qw \\
 \lstick{\ket{0}_k}  & & \gate{H} & \ctrl{-1}& \qw \\
 \lstick{\ket{0}_j}  & & \gate{H} & \ctrl{1}  & \qw  \\
\lstick{\ket{0}_i}  & & \gate{H} & \ctrl{-1} & \qw 
                      }$$

\item

we can reach a state in $V_{ijkl}$ as follows,

$$
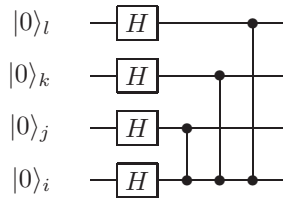
\Qcircuit @C=1em @R=.7em {  \lstick{\ket{0}_l}  & & \gate{H} & \qw& \qw &\ctrl{1}&\qw\\
 \lstick{\ket{0}_k}  & & \gate{H} & \qw& \ctrl{1}&\qw&\qw \\
 \lstick{\ket{0}_j}  & & \gate{H} & \ctrl{1}  & \qw &\qw&\qw \\
\lstick{\ket{0}_i}  & & \gate{H} & \ctrl{-1} & \ctrl{-1} &\ctrl{-3}&\qw
                      }$$

\item

we can reach a state in $V_{ij/kl}$ as follows, 

$$
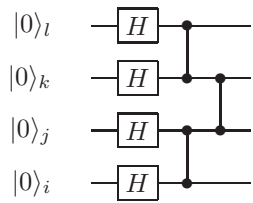
\Qcircuit @C=1em @R=.7em {  \lstick{\ket{0}_l}  & & \gate{H} & \ctrl{1}& \qw &\qw\\
 \lstick{\ket{0}_k}  & & \gate{H} & \ctrl{-1}& \ctrl{1}&\qw \\
 \lstick{\ket{0}_j}  & & \gate{H} & \ctrl{1}  & \ctrl{-1} &\qw \\
\lstick{\ket{0}_i}  & & \gate{H} & \ctrl{-1} & \qw &\qw
                      }$$
                      
  \item
we can reach a state in $V_{lr}$ as follows, 

$$
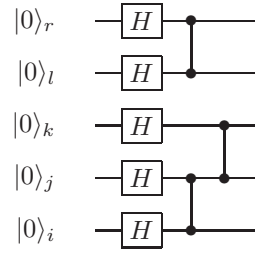
\Qcircuit @C=1em @R=.7em {  \lstick{\ket{0}_r}  & & \gate{H} & \ctrl{1}& \qw &\qw\\
 \lstick{\ket{0}_l}  & & \gate{H} & \ctrl{-1}& \qw &\qw\\
 \lstick{\ket{0}_k}  & & \gate{H} & \qw& \ctrl{1}&\qw \\
 \lstick{\ket{0}_j}  & & \gate{H} & \ctrl{1}  & \ctrl{-1} &\qw \\
\lstick{\ket{0}_i}  & & \gate{H} & \ctrl{-1} & \qw &\qw
                      }$$
                                                                              
    \item

we can reach a state in $W_0$ as follows, 

$$
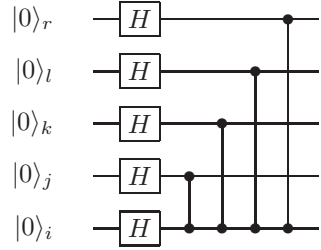
\Qcircuit @C=1em @R=.7em {    \lstick{\ket{0}_r}  & & \gate{H} & \qw& \qw &\qw&\ctrl{1}&\qw\\
\lstick{\ket{0}_l}  & & \gate{H} & \qw& \qw &\ctrl{1}&\qw&\qw\\
 \lstick{\ket{0}_k}  & & \gate{H} & \qw& \ctrl{1}&\qw&\qw&\qw \\
 \lstick{\ket{0}_j}  & & \gate{H} & \ctrl{1}  & \qw &\qw&\qw&\qw \\
\lstick{\ket{0}_i}  & & \gate{H} & \ctrl{-1} & \ctrl{-1} &\ctrl{-3}&\ctrl{-4}&\qw
                      }$$

  \item

we can reach a state in $W_{lr}$ as follows,

$$
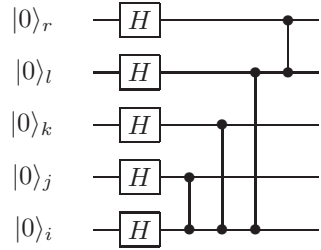
\Qcircuit @C=1em @R=.7em {    \lstick{\ket{0}_r}  & & \gate{H} & \qw& \qw &\qw&\ctrl{1}&\qw\\
\lstick{\ket{0}_l}  & & \gate{H} & \qw& \qw &\ctrl{1}&\ctrl{-1}&\qw\\
 \lstick{\ket{0}_k}  & & \gate{H} & \qw& \ctrl{1}&\qw&\qw&\qw \\
 \lstick{\ket{0}_j}  & & \gate{H} & \ctrl{1}  & \qw &\qw&\qw&\qw \\
\lstick{\ket{0}_i}  & & \gate{H} & \ctrl{-1} & \ctrl{-1} &\ctrl{-3}&\qw&\qw
                      }$$
\item

we can reach a state in $W_{jk/lr}$ as follows,

$$
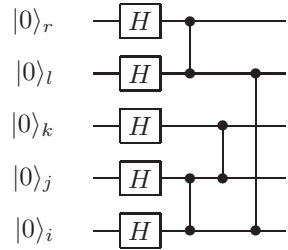
\Qcircuit @C=1em @R=.7em {   
\lstick{\ket{0}_r}  & & \gate{H} & \ctrl{1}  & \qw       &\qw      &\qw\\
\lstick{\ket{0}_l}  & & \gate{H} & \ctrl{-1} & \qw      &\ctrl{1}  &\qw\\
\lstick{\ket{0}_k} & & \gate{H} & \qw      & \ctrl{1}  &\qw      &\qw \\
\lstick{\ket{0}_j}  & & \gate{H} & \ctrl{1}  & \ctrl{-1} &\qw      &\qw \\
\lstick{\ket{0}_i}  & & \gate{H} & \ctrl{-1} & \qw       &\ctrl{-3}&\qw
                      }$$
\item

we can reach a state in $X_0$ as follows,

$$
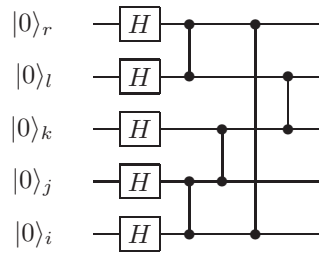
\Qcircuit @C=1em @R=.7em {   
\lstick{\ket{0}_r}  & & \gate{H} & \ctrl{1}  & \qw      &\ctrl{1}     &\qw &\qw\\
\lstick{\ket{0}_l}  & & \gate{H} & \ctrl{-1} & \qw      &\qw        &\ctrl{1} &\qw\\
\lstick{\ket{0}_k} & & \gate{H} & \qw      & \ctrl{1}  &\qw       &\ctrl{-1} &\qw\\
\lstick{\ket{0}_j}  & & \gate{H} & \ctrl{1}  & \ctrl{-1} &\qw      &\qw &\qw\\
\lstick{\ket{0}_i}  & & \gate{H} & \ctrl{-1} & \qw       &\ctrl{-4}&\qw &\qw
                      }$$
                
\end{itemize}

\section{Main results for Clifford states with real entries}

We are also interested in the real Clifford 5-qubit states. Even the study of complex states has garnered more attention than that of real ones, studying the real states is interesting in many ways. Each one of the 93 complex orbits contains at least one real state, and the real orbits are more numerous than the complex ones, and their structure is more intricate as we will see; however, each of the orbits is much smaller, with the largest orbits having size 4096. This makes the real orbits much easier to compute, and allows us to use them as a stepping stone toward the calculation of the complex orbits via the phase gate. We can define an equivalence relation on the real 5-qubit states, with two states being related if and only if there is a real local Clifford gate that takes one state to the other. Since not every pair of states that are related through a complex Clifford gate is also related through a real Clifford gate, it follows that, for each of the 93 complex orbits, the real states form one or more real orbits. Calculation using Mathematica shows that:
\begin{itemize}
\item The orbit $S_0$ contains one real orbit, $S_0^r;$
\item Each of the 10 orbits $T_{ij}$ contains one real orbit $T_{ij}^r;$
\item Each of the 10 orbits $U_{ijk}$ contains two real orbits, $U_{ijk}^r$ and $\ha U_{ijk}^r;$
\item Each of the 15 orbits $U_{ij/kl}$ contains one real orbit, $U_{ij/kl}^r;$
\item Each of the 5 orbits $V_{ijkl}$ contains two real orbits, $V_{ijkl}^r$ and $\ha V_{ijkl}^r;$
\item Each of the 15 orbits $V_{ij/kl}$ contains three real orbits, $V_{ij/kl}^r, \ha V_{ij,\bar{ijkl}}^r,$ and $\ha V_{kl,\bar{ijkl}}^r;$
\item Each of the 10 orbits $V_{ij}$ contains two real orbits, $V_{ij}^r$ and  $\ha V_{ij}^r;$
\item The orbit $W_0$ contains two real orbits, $W_0^r$ and $\ha W_0^r;$
\item Each of the 10 orbits $W_{ij}$ contains three real orbits, $W_{ij}^r, \ha W_{ij}^r,$ and $\ti W_{ij}^r;$
\item Each of the 15 orbits $W_{ij/kl}$ contains five real orbits, $W_{ij/kl}^r, \ha W_{ij/kl}^r, \ha W_{ij,\bar{ijkl}}^r, \ha W_{kl,\bar{ijkl}}^r,$ and $\ti W_{ij/kl}^r;$
\item The orbit $X_0$ contains 12 real orbits, as described below. 
\end{itemize}

The 12 real orbits contained in $X_0$ are labeled $X_{\left<ijklm\right>}^r$, where $\left<ijklm\right>$ denotes a bidirectional cycle on $\{1, 2, 3, 4, 5\}.$ Bidirectional cycles are permutations of $\{1,2,3,4,5\},$ under the equivalence relation that equates $\left<ijkls\right>$ with both $\left<jklsi\right>$ and $\left<slkji\right>.$ The number of distinct cycles on $\{1,2,3,4,5\}$ is $4! = 24,$ but the bidirectional nature of these cycles halves their number to 12. The real orbits are labeled using bidirectional cycles because these labels allows one to accurately predict the action of real Clifford gates on the various orbits.

In addition, another new type of labels is used for the real orbits that was not used for the complex ones; it is of the form $ij,k.$ The numbers $i$ and $j$ are interchangeable, but $k$ cannot be interchanged with $i$ or $j$ without changing the orbit. The number of possible subscripts of this form is ${5 \choose 2} \cdot 3 = 30.$ Thus there are two types of real orbits that are represented by 30 different orbits, namely, the $\ha V_{ij,k}$ and the $\ha W_{ij,k}.$ For instance, the complex orbit $V_{12/34}$ contains two such real orbits, $\ha V_{12,5}$ and $\ha V_{34,5}.$

Table \ref{table4} shows the size of each real orbit, as well as the number of orbits of each type.

\begin{table}[h]
\setlength\extrarowheight{1mm}
\begin{tabular}{|c|c|c||c|c|c|} %the | are vertical bars, not lowercase Ls
\hline
class  & size & no. of & class & size & no. of \\
type & & orbits & type & & orbits \\
\hline
$S_0^r$ & $2^{11} = 2048$ & 1 & $W_0^r$ & $2^{11} = 2048$ & 1 \\
\hline 
$T_{ij}^r$ & $2^{10} = 1024$ & 10 & $\ha W_0^r$ & $2^{7} = 128$ & 1 \\
\hline 
$U_{ijk}^r$ & $2^{11} = 2048$ & 10 & $W_{ij}^r$ & $2^{11}=2048$ & 10 \\
\hline
$\ha U_{ijk}^r$ & $2^9 = 512$ & 10 & $\ha W_{ij}^r$ & $2^{10}=1024$ & 10 \\
\hline
$U_{ij/kl}^r$ & $2^9 = 512$ & 15 & $\ti W_{ij}^r$ & $2^{9} = 512$ & 10 \\
\hline 
$V_{ijkl}^r$ & $2^{11} = 2048$ & 5 & $W_{ij/kl}^r$ & $2^{11}=2048$ & 15 \\
\hline
$\ha V_{ijkl}^r$ & $2^{8} = 256$ & 5 & $\ha W_{ij/kl}^r$ & $2^{11}=2048$ & 15\\
\hline
$V_{ij/kl}^r$ & $2^{11} = 2048$ & 15 & $\ha W_{ij,k}^r$ & $2^{10} = 1024$ & 30 \\
\hline
$\ha V_{ij,k}^r$ & $2^{10} = 1024$ & 30 & $\ti W_{ij/kl}^r$ & $2^{9} = 512$ & 15 \\
\hline 
$V_{ij}^r$ & $2^{10} = 1024$ & 10 & $X_{\left<ijkls\right>}^r$ & $2^{11}=2048$ & 12 \\
\hline 
$\ha V_{ij}^r$ & $2^{8} = 256$ & 10 &&& \\
\hline
\end{tabular}
\caption{Size of the 21 types of orbits for real Clifford states.} \label{table4}
\end{table}

We see from the table that the total number of orbits is 240, the total number of orbit types is 21, and there are 84 orbits of size 2048, 90 orbits of size 1024, 50 orbits of size 512, 15 orbits of size 256, and one orbit of size 128, for a total of 293760 5-qubit real Clifford states. We see that this is identical to the number of 4-qubit complex Clifford states. \cite{LP}

The action of each CZ gate on each real orbit is summarized in Tables \ref{tablerA} and \ref{tablerB}, which due to their sizes are included in the appendix. The graph in Figure \ref{fig2} shows the relationship between the 240 orbits; however, in order to make the graph readable, only one of the orbits of each type is displayed. The presence of an edge between two nodes indicates the existence of a CZ gate and of two orbits, one of each type, taking a state in one orbit to a state in the other orbit. Since the CZ gate is involutory, all edges are bidirectional. We see that each real Clifford state can be reached from the unentangled states using a maximum of 7 CZ gates, together with the Hadamard gate and Z-gate. There is only one orbit that requires 7 CZ gates, and that orbit is the single orbit of size 128 (the smallest orbit). Interestingly, that orbit is {\em not} contained in the complex orbit $X_0$ that was located at distance 5 from $S_0$ in Figure \ref{fig1}; rather, it is contained in $W_0,$ which is at distance 4 from $S_0$ in Figure \ref{fig1}.

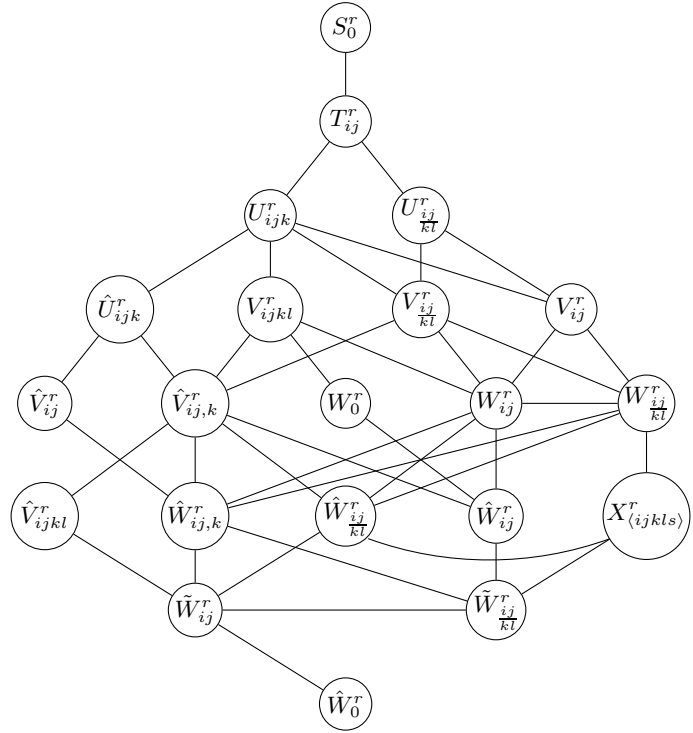
\begin{figure}
\begin{center}
    \begin{tikzpicture}
        \node at (0,0) [circleA] (A) {$S_0^r$};
        \node at (0,-1.25) [circleA] (B) {$T_{ij}^r$};
        \node at (-1,-2.5) [circleA] (C) {$U_{ijk}^r$};
        \node at (1,-2.5) [circleA] (D) {$U_{\frac{ij}{kl}}^r$};
        \node at (-3,-3.75) [circleB] (Cx) {$\ha U_{ijk}^r$};
        \node at (-1,-3.75) [circleB] (E) {$V_{ijkl}^r$};
        \node at (1,-3.75) [circleA] (F) {$V_{\frac{ij}{kl}}^r$};
        \node at (3,-3.75) [circleA] (G) {$V_{ij}^r$};
        \node at (-4,-5) [circleA] (Gx) {$\ha V_{ij}^r$};
        \node at (-2,-5) [circleB] (Fx) {$\ha V_{ij,k}^r$};
        \node at (0,-5) [circleA] (H) {$W_0^r$};
        \node at (2,-5) [circleA] (I) {$W_{ij}^r$};
        \node at (4,-5) [circleA] (J) {$W_{\frac{ij}{kl}}^r$};
        \node at (2,-6.5) [circleA] (Hx) {$\ha W_{ij}^r$};
        \node at (-4,-6.5) [circleB] (Ex) {$\ha V_{ijkl}^r$};
        \node at (0,-6.5) [circleA] (Jx) {$\ha W_{\frac{ij}{kl}}^r$};
        \node at (-2,-6.5) [circleB] (Jxx) {$\ha W_{ij,k}^r$};
        \node at (4,-6.5) [circleC] (K) {$X_{\left<ijkls\right>}^r$};
        \node at (2,-7.75) [circleA] (L) {$\ti W_{\frac{ij}{kl}}^r$};
        \node at (-2,-7.75) [circleA] (M) {$\ti W_{ij}^r$};
        \node at (0,-9) [circleA] (Z) {$\ha W_0^r$};
        \draw (A) -- (B);
        \draw (B) -- (C); 
        \draw (B) -- (D);
        \draw (C) -- (E); 
        \draw (C) -- (F);
        \draw (C) -- (G);
        \draw (C) -- (Cx);
        \draw (D) -- (F); 
        \draw (D) -- (G);
        \draw (Cx) -- (Fx); 
        \draw (Cx) -- (Gx); 
        \draw (E) -- (H); 
        \draw (E) -- (I);
        \draw (E) -- (Fx);
        \draw (F) -- (I); 
        \draw (F) -- (J);
        \draw (F) -- (Fx);
        \draw (G) -- (I); 
        \draw (G) -- (J);
        \draw (H) -- (Hx);
        \draw (I) -- (Hx); 
%        \draw (J) -- (Hx);
        \draw (Fx) -- (Hx); 
        \draw (Fx) -- (Ex);
        \draw (I) -- (Jx); 
        \draw (J) -- (Jx);
        \draw (Gx) -- (Jxx);
        \draw (I) -- (Jxx); 
        \draw (J) -- (Jxx);
        \draw (Fx) -- (Jx);
        \draw (Fx) -- (Jxx);
        \draw (I) -- (J);
        \draw (J) -- (K);
        \draw (Hx) -- (L);
        \draw (Jxx) -- (L);
        \draw (K) -- (L);
        \draw (Ex) -- (M);
        \draw (Jxx) -- (M);
        \draw (Jx) -- (M);
        \draw (L) -- (M);
        \draw (M) -- (Z);
        \draw (0.3,-6.8) arc ( 250:290:4.7 );
    \end{tikzpicture}
    \end{center}
        \caption{Actions of CZ gates on orbits of real Clifford states.}  \label{fig2}
\end{figure}

\section{Conclusions} 

\begin{itemize}

\item

The work in this paper can be viewed as a continuation of the papers \cite{P} for Clifford 3-qubit states and \cite{LP} for Clifford 4-qubit states. Even though we have made progress for the case of 6 qubits, we still do not have a classification. This study shows how entanglement happens when we use the non-local gate CZ. We have visualized some general properties but not enough to get a description of the Clifford $n$-qubit states in general. 

\item

Markus Grassl has inform us that, for Clifford 6 and 7 qubits, Danielsen and Parker, \cite{L}, have found 26, for 6 qubits, and 59, for 7 qubits, equivalence classes with respect to the Clifford group and qubit permutation. Dr. Grassl also have told us that using the 26 equivalence classes for 6 qubits, he has found 760 different classes excluding qubit permutations and using the 59 for 7 qubits, he has found 10773 different classes excluding qubit permutations.

\item
Recall that the support of a state $\ket{\phi}$ is the number of nonzero coefficients when $\ket{\phi}$ is written in the computational basis, \cite{D}. We have that there are 1990656  Clifford 5-qubit absolutely maximally entangled (AME) States. 1081344 have support 32, 786432 have support 16 and 122880 have support 8. We have that all the AME states are in one orbit, this is, they differ by local Clifford gates. Out of the 1990656 AME states, 24576 have real amplitudes. This time not all of them differ by local gates generated by the Hadamard gate and Z-gate. There are 12 orbits, each one with 2048 AME 5-qubit states. Each orbit have the property that two states in the same orbit differ by a local gate generated by the Hadamard gate and Z-gate.

\item
For the 5-qubit case, any pair of Clifford states can be connected using local gates and at most 5 CZ gates. Any pair of Clifford states with real amplitudes can be connected using local Clifford gates with real entries and at most 7 CZ gates. We conjecture that, for the $n$-qubit case ($n \geq 2$), any pair of Clifford states with real amplitudes can be connected using local Clifford gates with real entries and at most $2n-3$ CZ gates. We have verified this to be true up to $n=6.$

\end{itemize}

{\bf Acknowledgments}

The authors would like to express their gratitude to Markus Grassl, for providing many helpful comments and data, and to William Munizzi, for a very helpful discussion of this paper and of related work. \vspace{5mm}

\section{Declarations}

{\bf Conflict of interest}

The authors declare no competing interests.

{\bf Contributions}

All authors contributed equally to the paper.

{\bf Funding}

Oscar Perdomo was partially supported by a AAUP-CSU research project at Central Connecticut State University.

{\bf Data availability}

The datasets generated during and/or analyzed during the current study are available upon further request.

\vspace{13cm}
 
 \section*{Appendix: Data tables}
 
 The tables in this appendix show in detail the action of each of the CZ gate on each of the orbits, for both the Clifford 5-qubit case and the real Clifford 5-qubit case.
 
\begin{table}[H]
\setlength\extrarowheight{1mm}
\begin{center}
\begin{tabular}{|c|c|c|} %the | are vertical bars, not lowercase Ls
\hline
orbit type & CZ & number of states  \\
&  applied&  mapped to each orbit \\
\hline
$S_0$ & ${ij}$ & 34560 to $S_0$ \\
& & 27648 to $T_{ij}$ \\
\hline 
$T_{ij}$ & ${ij}$ & 13824 to $T_{ij}$ \\
& & 27648 to $S_0$ \\
\hline 
$T_{ij}$ & ${ik}$ & 13824 to $T_{ij}$ \\
&  & 27648 to $U_{ijk}$ \\
\hline 
$T_{ij}$ & ${kl}$ & 23040 to $T_{ij}$ \\
&  & 18432 to $U_{ij/kl}$ \\
\hline 
$U_{ijk}$ & ${ij}$ & 69120 to $U_{ijk}$ \\
& & 27648 to $T_{ik}$ \\
& & 27648 to $T_{jk}$ \\
\hline 
$U_{ijk}$ & ${il}$ & 41472 to $U_{ijk}$ \\
& & 27648 to $V_{ijkl}$ \\
& & 55296 to $V_{il/jk}$ \\
\hline 
$U_{ijk}$ & ${ls}$ & 69120 to $U_{ijk}$ \\
& & 55296 to $V_{ls}$ \\
\hline %Startng the Tij,kl
$U_{ij/kl}$ & ${ij}$ & 9216 to $U_{ij/kl}$ \\
& & 18432 to $T_{kl}$ \\
\hline
$U_{ij/kl}$ & ${ik}$ & 27648 to $V_{ij/kl}$ \\
\hline
$U_{ij/kl}$ & ${is}$ & 9216 to $U_{ij/kl}$ \\
& & 18432 to $V_{kl}$ \\
\hline%Startng the Tijkl
$V_{ijkl}$ & ${ij}$ & 13824 to $V_{ijkl}$ \\
& & 55296 to $V_{ij/kl}$ \\
& & 27648 to $U_{jkl}$ \\
& & 27648 to $U_{ikl}$ \\
\hline
$V_{ijkl}$ & ${is}$ & 41472 to $V_{ijkl}$ \\
& & 27648 to $W_0$ \\
& & 55296 to $W_{is}$ \\
\hline %Startng the Xij,kl
$V_{ij/kl}$ & ${ij}$ & 82944 to $V_{ij/kl}$ \\
& & 55296 to $V_{ijkl}$ \\
& & 55296 to $U_{jkl}$ \\
& & 55296 to $U_{ikl}$ \\
\hline
$V_{ij/kl}$ & ${ik}$ & 110592 to $V_{ij/kl}$ \\
& & 27648 to $U_{ij/kl}$ \\
& & 110592 to $V_{il/jk}$ \\
\hline%czis
$V_{ij/kl}$ & ${is}$ &82944 to $V_{ij/kl}$ \\
& & 55296 to $W_{kl}$ \\
& & 110592 to $W_{is/kl}$ \\
\hline
\end{tabular}
\end{center}
\caption{Action of CZ gates on orbits of Clifford states, part 1} \label{table2}
\end{table}

\begin{table}[H]
\setlength\extrarowheight{1mm}
\begin{center}
\begin{tabular}{|c|c|c|} %the | are vertical bars, not lowercase Ls
\hline
orbit type & CZ & number of states  \\
&  applied&  mapped to each orbit \\
\hline
$V_{ij}$ & ${ij}$ & 27648 to $V_{ij}$ \\
& & 55296 to $U_{\bar{ij}}$ \\
\hline
$V_{ij}$ & ${ik}$ & 27648 to $W_{ij}$ \\
& & 55296 to $W_{ij/\bar{ijk}}$ \\
\hline 
$V_{ij}$ & ${kl}$ & 46080 to $V_{ij}$ \\
&  & 18432 to $U_{ij/\bar{ijk}}$ \\
&  & 18432 to $U_{ij/\bar{ijl}}$ \\
\hline 
$W_0$ & ${ij}$ & 13824 to $W_0$ \\
&  & 55296 to $W_{ij}$ \\
&  & 27648 to $V_{\bar{i}}$ \\
&  & 27648 to $V_{\bar{j}}$ \\
\hline 
$W_{ij}$ & ${ij}$ & 82944 to $W_{ij}$ \\
& & 55296 to $W_0$ \\
& & 55296 to $V_{\bar{i}}$ \\
& & 55296 to $V_{\bar{j}}$ \\
\hline
$W_{ij}$ & ${ik}$ & 55296 to $W_{ij}$ \\
& & 27648 to $V_{ij}$ \\
& & 110592 to $W_{jk/\bar{ijk}}$ \\
& & 55296 to $W_{ij/\bar{ijk}}$ \\
\hline 
$W_{ij}$ & ${kl}$ & 27648 to $W_{ij}$ \\
& & 55296 to $V_{ij/\bar{ijk}}$ \\
& & 55296 to $V_{ij/\bar{ijl}}$ \\
& & 110592 to $W_{ij/kl}$ \\
\hline
$W_{ij/kl}$ & ${ij}$ & 165888 to $W_{ij/kl}$ \\
& & 110592 to $W_{kl}$ \\
& & 110592 to $V_{kl/\bar{ikl}}$ \\
& & 110592 to $V_{kl/\bar{jkl}}$ \\
& & 27648 to $U_{ikl}$ \\
\hline
$W_{ij/kl}$ & ${ik}$ & 55296 to $W_{ij/kl}$ \\
& & 110592 to $W_{kl/\bar{jkl}}$ \\
& & 110592 to $W_{ij/\bar{ijl}}$ \\
& & 221184 to $X_0$ \\
\hline
$W_{ij/kl}$ & ${is}$ & 165888 to $W_{ij/kl}$ \\
& & 110592 to $W_{js/kl}$ \\
& & 110592 to $W_{js}$ \\
& & 55296 to $W_{ij}$ \\
& & 55296 to $V_{ij}$ \\
\hline
$X_0$ & ${ij}$ & 663552 to $X_0$ \\
& & 221184 to $W_{ia/jb}$ \\
& & for each  $(a,b),$ \\ 
& & $a,b \in \bar{ij}, a\neq b$ \\
& & (there are 6 such \\
& & $(a,b)$ for each $ij$) \\
\hline
\end{tabular}
\end{center}
\caption{Action of CZ gates on orbits of Clifford states, part 2} \label{table3}
\end{table}

\begin{table}[H]
\setlength\extrarowheight{0.35mm}
\begin{center}
\begin{tabular}{|c|c|c||c|c|c|c|} %the | are vertical bars, not lowercase Ls
\hline
orbit & CZ & {\tiny number of states} & orbit & CZ & {\tiny number of states} \\
type &  {\tiny{applied}}&  {\tiny states mapped} & type &  {\tiny{applied}}&  {\tiny states mapped} \\
&& {\tiny to each orbit} & && {\tiny to each orbit} \\
\hline
$S_0^r$ & ${ij}$ & 1536 to $S_0^r$ & $V_{ij/kl}^r$ & ${ij}$ & 512 to $V_{ij/kl}^r$ \\
& & 512 to $T_{ij}^r$ & & & 512 to $\ha V_{ij,\bar{ijkl}}^r$ \\
& &&& & 512 to $U_{jkl}^r$ \\
&& & & & 512 to $U_{ikl}^r$ \\
\hline 
$T_{ij}^r$ & ${ij}$ & 512 to $T_{ij}^r$ & $V_{ij/kl}^r$ & ${ik}$ & 512 to $V_{il/jk}^r$ \\
& & 512 to $S_0^r$ & & & 512 to $U_{ij/kl}^r$ \\
&&&& & 512 to $\ha V_{ij,\bar{ijkl}}^r$ \\
&&&& & 512 to $\ha V_{kl,\bar{ijkl}}^r$ \\
\hline 
$T_{ij}^r$ & ${ik}$ & 512 to $T_{ij}^r$ & $V_{ij/kl}^r$ & ${is}$ & 1024 to $V_{ij/kl}^r$ \\
&& 512 to $U_{ijk}^r$ & & & 512 to $W_{kl}^r$ \\
&&&& & 512 to $W_{is/kl}^r$ \\
\hline 
$T_{ij}^r$ & ${kl}$ & 768 to $T_{ij}^r$ & $\ha V_{ij,k}^r$ & ${ij}$ & 512 to $V_{ij/\bar{ijk}}^r$ \\
&  & 256 to $U_{ij/kl}^r$ & & & 512 to $V_{\bar{k}}^r$ \\
\hline 
$U_{ijk}^r$ & ${ij}$ & 512 to $U_{ijk}^r$ & $\ha V_{ij,k}^r$ & ${ik}$ & 512 to $V_{ij,k}^r$ \\
& & 512 to $\ha U_{ijk}^r$ &  &  & 512 to $\ha W_{ik/\bar{ijk}}^r$ \\
& & 512 to $T_{ik}^r$ &&&\\
& & 512 to $T_{jk}^r$ &&&\\
\hline 
$U_{ijk}^r$ & ${il}$ & 1024 to $U_{ijk}^r$ & $\ha V_{ij,k}^r$ & ${il}$ & 512 to $V_{ij/\bar{ijk}}^r$ \\
& & 512 to $V_{ijkl}^r$ &  &  & 512 to $\ha V_{\bar{jkl},k}^r$ \\
& & 512 to $V_{il/jk}^r$ &&&\\
\hline 
$U_{ijk}^r$ & ${ls}$ & 1536 to $U_{ijk}^r$ & $\ha V_{ij,k}^r$ & ${kl}$ & 512 to $\ha V_{ij,k}^r$ \\
& & 512 to $V_{ls}^r$ &  &  & 256 to $\ha W_{kl,\bar{ijkl}}^r$ \\
&  &  &  &  & 256 to $\ha W_{ij}^r$ \\
\hline 
$\ha U_{ijk}^r$ & ${ij}$ & 512 to $U_{ijk}^r$ & $\ha V_{ij,k}^r$ & ${ls}$ & 256 to $\ha V_{ij,k}^r$ \\
& &  &  &  & 256 to $\ha V_{ijls}^r$ \\
& &  &  &  & 256 to $\ha U_{ijl}^r$ \\
& &  &  &  & 256 to $\ha U_{ijs}^r$ \\
\hline 
$\ha U_{ijk}^r$ & ${il}$ & 256 to $\ha U_{ijk}^r$ & $V_{ij}^r$ & ${ij}$ & 512 to $V_{ij}^r$ \\
& & 256 to $\ha V_{ij,\bar{ijkl}}^r$ &  &  & 512 to $U_{\bar{ij}}^r$ \\
\hline 
$\ha U_{ijk}^r$ & ${ls}$ & 384 to $\ha U_{ijk}^r$ & $V_{ij}^r$ & ${ik}$ & 512 to $W_{ij}^r$ \\
& & 128 to $\ha V_{ls}^r$ &  &  & 512 to $W_{ij/\bar{ijk}}^r$ \\
\hline %Startng the Tij,kl
$U_{ij/kl}^r$ & ${ij}$ & 256 to $U_{ij/kl}^r$ & $V_{ij}^r$ & ${kl}$ & 256 to $V_{ij}^r$ \\
& & 256 to $T_{kl}^r$ &  &  & 256 to $\ha V_{ij}^r$ \\
&  &  &  &  & 256 to $U_{ij/\bar{ijk}}^r$ \\
&  &  &  &  & 256 to $U_{ij/\bar{ijl}}^r$ \\
\hline
$U_{ij/kl}^r$ & ${ik}$ & 512 to $V_{ij/kl}^r$ & $\ha V_{ij}^r$ & ${ij}$ & 128 to $\ha V_{ij}^r$ \\
&&&&& 128 to $\ha U_{\bar{ij}}^r$ \\
\hline
$U_{ij/kl}^r$ & ${is}$ & 256 to $U_{ij/kl}^r$ & $\ha V_{ij}^r$ & ${ik}$ & 256 to $\ha W_{ij,k}^r$ \\
& & 256 to $V_{kl}^r$ &&&\\
\hline%Startng the Tijkl
$V_{ijkl}^r$ & ${ij}$ & 512 to $V_{ijkl}^r$ & $\ha V_{ij}^r$ & ${kl}$ & 256 to $V_{ij}^r$ \\
& & 512 to $\ha V_{kl,\bar{ijkl}}^r$ &&&\\
& & 512 to $U_{jkl}^r$ &&&\\
& & 512 to $U_{ikl}^r$ &&&\\
\hline
$V_{ijkl}^r$ & ${is}$ & 1024 to $V_{ijkl}^r$ & $W_0^r$ & ${ij}$ & 512 to $W_0^r$ \\
& & 512 to $W_0^r$ &&& 512 to $V_{\bar{i}}^r$ \\  
& & 512 to $W_{is}^r$ &&& 512 to $V_{\bar{j}}^r$ \\
&  &  &  &  & 512 to $\ha W_{ij}^r$ \\
\hline%Startng the Tijkl
$\ha V_{ijkl}^r$ & ${ij}$ & 256 to $\ha V_{kl,\bar{ijkl}}^r$ & $\ha W_0^r$ & ${ij}$ & 128 to $\ti W_{ij}^r$ \\
\hline
$\ha V_{ijkl}^r$ & ${is}$ & 128 to $\ha V_{ijkl}^r$ &&&\\
& & 128 to $\ti W_{is}^r$ &&&\\
\hline
\end{tabular}
\end{center}
\caption{Action of CZ gates on orbits of real Clifford states, part 1} \label{tablerA}
\end{table}

\begin{table}[H]
\setlength\extrarowheight{0.35mm}
\begin{center}
\begin{tabular}{|c|c|c||c|c|c|c|} %the | are vertical bars, not lowercase Ls
\hline
orbit & CZ & {\tiny number of states} & orbit & CZ & {\tiny number of states} \\
type &  {\tiny{applied}}&  {\tiny states mapped} & type &  {\tiny{applied}}&  {\tiny states mapped} \\
&& {\tiny to each orbit} & && {\tiny to each orbit} \\
\hline
$W_{ij}^r$ & ${ij}$ & 512 to $W_{ij}^r$ & $\ha W_{ij/kl}^r$ & ${ij}$ & 512 to $\ha W_{ij/kl}^r$ \\
& & 512 to $\ha W_{ij}^r$ &  & & 512 to $\ti W_{kl}^r$ \\
& & 512 to $V_{\bar{i}}^r$ &&& 512 to $\ha V_{js,i}^r$ \\
&&  512 to $V_{\bar{j}}^r$ &&& 512 to $\ha V_{is,j}^r$ \\
\hline 
$W_{ij}^r$ & ${ik}$ & 512 to $V_{ij}^r$ & $\ha W_{ij/kl}^r$ & ${ik}$ & 512 to $W_{ij/kl}^r$ \\
& & 512 to $\ha W_{ij}^r$ &  & & 512 to $\ha W_{is/kl}^r$ \\
&& 512 to $W_{jk/\bar{ijk}}^r$ &  & & 512 to $\ha W_{ij/ks}^r$ \\
&& 512 to $\ha W_{ij/\bar{ijk}}^r$ &  & & 512 to $X_{\left<kjsli\right>}^r$ \\
\hline 
$W_{ij}^r$ & ${kl}$ & 512 to $W_{ij}^r$ & $\ha W_{ij/kl}^r$ & ${is}$ & 512 to $W_{ij/kl}^r$ \\
& & 512 to $V_{ij/\bar{ijk}}^r$ & & & 512 to $W_{ij}^r$ \\
&& 512 to $V_{ij/\bar{ijl}}^r$ &&& 512 to $\ha W_{js}^r$ \\
&& 512 to $\ha W_{ij,\bar{ijkl}}^r$ &&& 512 to $\ha W_{kl,i}^r$ \\
\hline
$\ha W_{ij}^r$ & ${ij}$ & 512 to $W_{ij}^r$& $\ha W_{ij,k}^r$ & ${ij}$ & 256 to $\ha W_{ij,k}^r$ \\
& & 512 to $W_0^r$ & & & 256 to $\ha V_{\bar{ijk},i}^r$ \\
&&& & & 256 to $\ha V_{\bar{ijk},j}^r$ \\
&&& & & 256 to $\ti W_{ij/\bar{ijk}}^r$ \\
\hline 
$\ha W_{ij}^r$ & ${ik}$ & 512 to $W_{ij}^r$ & $\ha W_{ij,k}^r$ & ${ik}$ & 256 to $\ha V_{ij}^r$ \\
& & 512 to $\ha W_{jk/\bar{ijk}}^r$ & &&  256 to $\ti W_{ij}^r$ \\
&&& & & 256 to $\ha W_{jk,i}^r$ \\
&&& & & 256 to $\ti W_{ij/\bar{ijk}}^r$ \\
\hline 
$\ha W_{ij}^r$ & ${kl}$ & 256 to $\ha W_{ij}^r$ & $\ha W_{ij,k}^r$ & ${il}$ & 512 to $\ha W_{ij,s}^r$ \\
& & 256 to $\ti W_{ij/kl}^r$ & & & 512 to $X_{\left<ijlsk\right>}^r$ \\
&& 256 to $\ha V_{ij,k}^r$ &&& \\
&& 256 to $\ha V_{ij,l}^r$ &&& \\
\hline
$\ti W_{ij}^r$ & ${ij}$ & 128 to $\ti W_{ij}^r$ & $\ha W_{ij,k}^r$ & ${kl}$ & 512 to $W_{ij/ls}^r$ \\
& & 128 to $\ha V_{\bar{i}}^r$ & & & 512 to $\ha W_{ij/ks}^r$ \\
& & 128 to $\ha V_{\bar{j}}^r$ & & & \\
& & 128 to $\ha W_0^r$ & & & \\
\hline 
$\ti W_{ij}^r$ & ${ik}$ & 256 to $\ti W_{jk/\bar{ijk}}^r$ & $\ha W_{ij,k}^r$ & ${ls}$ & 512 to $W_{ij/ls}^r$ \\
& & 256 to $\ha W_{ij,k}^r$ & & & 512 to $W_{ij}^r$ \\
\hline 
$\ti W_{ij}^r$ & ${kl}$ & 512 to $\ha W_{ij/kl}^r$ & $\ti W_{ij/kl}^r$ & ${ik}$ & 256 to $\ha W_{kl}^r$ \\
& &&& & 256 to $\ha W_{ij,s}^r$ \\
\hline 
$W_{ij/kl}^r$ & ${ij}$ & 512 to $\ha W_{ij/kl}^r$ & $\ti W_{ij/kl}^r$ & ${is}$ & 512 to $X_{\left<jiskl\right>}^r$ \\
& & 512 to $V_{kl/\bar{ikl}}^r$ &  &  & \\
& & 512 to $V_{kl/\bar{jkl}}^r$ &  &  &  \\
& & 512 to $\ha W_{kl,\bar{ijkl}}^r$ &  &  &  \\
\hline 
$W_{ij/kl}^r$ & ${ik}$ & 512 to $W_{ij/\bar{ijl}}^r$ & $\ti W_{ij/kl}^r$ & ${is}$ & 256 to $\ti W_{js}^r$ \\
& & 512 to $W_{kl/\bar{jkl}}^r$ &&& 256 to $\ha W_{ij,s}^r$ \\
& & 512 to $\ha W_{ij/kl}^r$ &  &  &  \\
& & 512 to $X_{\left<ijslk\right>}^r$ &  &  &  \\
\hline 
$W_{ij/kl}^r$ & ${is}$ & 512 to $\ha W_{ij/kl}^r$ & $X_{\left<ijkls\right>}^r$ & ${ij}$ & 512 to $X_{\left<ijksl\right>}^r$ \\
& & 512 to $W_{js}^r$ &&& 512 to $X_{\left<ijlks\right>}^r$ \\
& & 512 to $V_{ij}^r$ &&& 512 to $W_{is/jk}^r$ \\
& & 512 to $\ha W_{kl,s}^r$ &&& 512 to $\ha W_{ik/js}$ \\
\hline
&&& $X_{\left<ijkls\right>}^r$ & ${ik}$ & 512 to $X_{\left<ijkls\right>}^r$ \\
&&&&& 512 to $\ti W_{is/kl}^r$ \\
&&&&& 512 to $\ha W_{ij,s}^r$ \\
&&&&& 512 to $\ha W_{jk,l}^r$ \\
\hline
\end{tabular}
\end{center}
\caption{Action of CZ gates on orbits of Clifford states, part 2} \label{tablerB}
\end{table}

\end{document}